\documentclass[pre,aps,superscriptaddress,twocolumn,floatfix]{revtex4-1}

\usepackage{dsfont}
\usepackage{mathtext}

\usepackage{mathtools}
\usepackage[usenames,dvipsnames]{xcolor}
\usepackage{amsmath}
\usepackage{amssymb,mathrsfs}
\usepackage{graphicx}
\usepackage{epstopdf}

\usepackage{mathtext}

\usepackage{bm}

\newcommand{\br}{{\bm r}}

\newcommand{\bs}{{\bm s}}

\begin{document}
	
	\title{Four-wave mixing Floquet topological soliton}

	\author{Sergey~K.~Ivanov}
	\affiliation{Moscow Institute of Physics and Technology, Institutsky lane 9, Dolgoprudny, Moscow region, 141700, Russia}
	\affiliation{Institute of Spectroscopy, Russian Academy of Sciences, Troitsk, Moscow, 108840, Russia}
	
	\author{Yaroslav~V.~Kartashov}
	\affiliation{Institute of Spectroscopy, Russian Academy of Sciences, Troitsk, Moscow, 108840, Russia}
	\affiliation{ICFO-Institut de Ciencies Fotoniques, The Barcelona Institute of Science and Technology, 08860 Castelldefels (Barcelona), Spain}
	
	\author{Vladimir~V.~Konotop}
	\affiliation{Departamento de F\'isica, Faculdade de Ci\^encias, Universidade de Lisboa, Campo Grande, Ed. C8, Lisboa 1749-016, Portugal}
	\affiliation{Centro de F\'isica Te\'orica e Computacional, Universidade de Lisboa, Campo Grande, Ed. C8, Lisboa 1749-016, Portugal}

\begin{abstract}
We consider a topological Floquet insulator realized as a honeycomb array of helical waveguides imprinted in weakly birefringent medium. The systems accounts for four-wave mixing occurring at a series of resonances arising due to Floquet phase matching. Under these resonant conditions the system sustains stable linearly polarized and metastable elliptically polarized two-component edge solitons. Coupled nonlinear equations describing evolution of the envelopes of such solitons are derived. 
\end{abstract}

\maketitle

Transverse periodic modulation of a nonlinear medium represents one of the most common tools for modification of the properties of a propagating beam. In particular, one can create artificial dispersion supporting diverse types of solitons and enhancing their stability, or create effective phase matching conditions for resonant wave interactions. Rich possibilities for light propagation control are offered by longitudinal periodic modulations of the medium~\cite{Rechtsman-13} allowing realization of Floquet topological insulators - one of the most interesting and important applications of such artificially created media \cite{linreview01,linreview02}.

A characteristic feature of topological insulators is the existence of states localized near their edges and propagating along them~\cite{HasanKane-10}. In nonlinear optical systems localized solitons can bifurcate from the linear topological edge states~\cite{SmirnovaLeykamChongKivshar-20}. A variety of solitons sustained by Floquet topological insulators was already obtained. These are 
discrete ~\cite{Lumer-13,Ablowitz-14,AblowitzCole-17} and continuous~\cite{LeykamChong-16,Ivanov-20} scalar edge solitons, two-component vector \cite{IvanovACS-20,IvanovDipol-21} and Bragg~\cite{IvanovBragg-20} edge solitons. Topological Floquet solitons were recently observed ~\cite{MukherjeeRechtsman-20} along with nonlinearity-induced edge states \cite{NonlinearTI}.

All multicomponent solitons mentioned above were obtained for non-resonant case, when components are coupled via cross-phase modulation only~\cite{ChenSegevCoskunChristodoulides-96}. At the same time, diverse optical systems with Kerr nonlinearity sustain vector solitons in the presence of resonances, involving the process of four-wave mixing (FWM)~\cite{Menyuk-fiber,Menyik-spatial,Sterke-91,Winful,ChenAtai-94}, see also \cite{Akhmediev1,Akhmediev2}. While FWM in 2D periodic media is well studied \cite{Bartal-06}, its effect on solitons in Floquet topological insulators has not been addressed, so far.

The goal of the present Letter is to report a new type of topological solitons -- FWM Floquet solitons. Such elliptically polarized vector states are composed of two orthogonally polarized components and propagate along the edge of photonic insulator, traversing hundreds of array periods with negligible radiative losses. In sharp contrast to non-Floquet media, the resonant conditions for the efficient FWM can be satisfied in our system for formally infinite number of propagation constants determined by the novel {\em Floquet matching conditions}. {In contrast to previously studied scalar topological solitons, the FWM Floquet solitons are essentially vector objects described by two coupled equations for slowly varying amplitudes averaged over the Floquet period. Their stability properties strongly depend on effective phase mismatch controlled by waveguide twisting. Propagation of such solitons is characterized by fast small-amplitude oscillations superimposed over the average amplitudes.} 

\begin{figure}[t]
\centering\includegraphics[width=\linewidth]{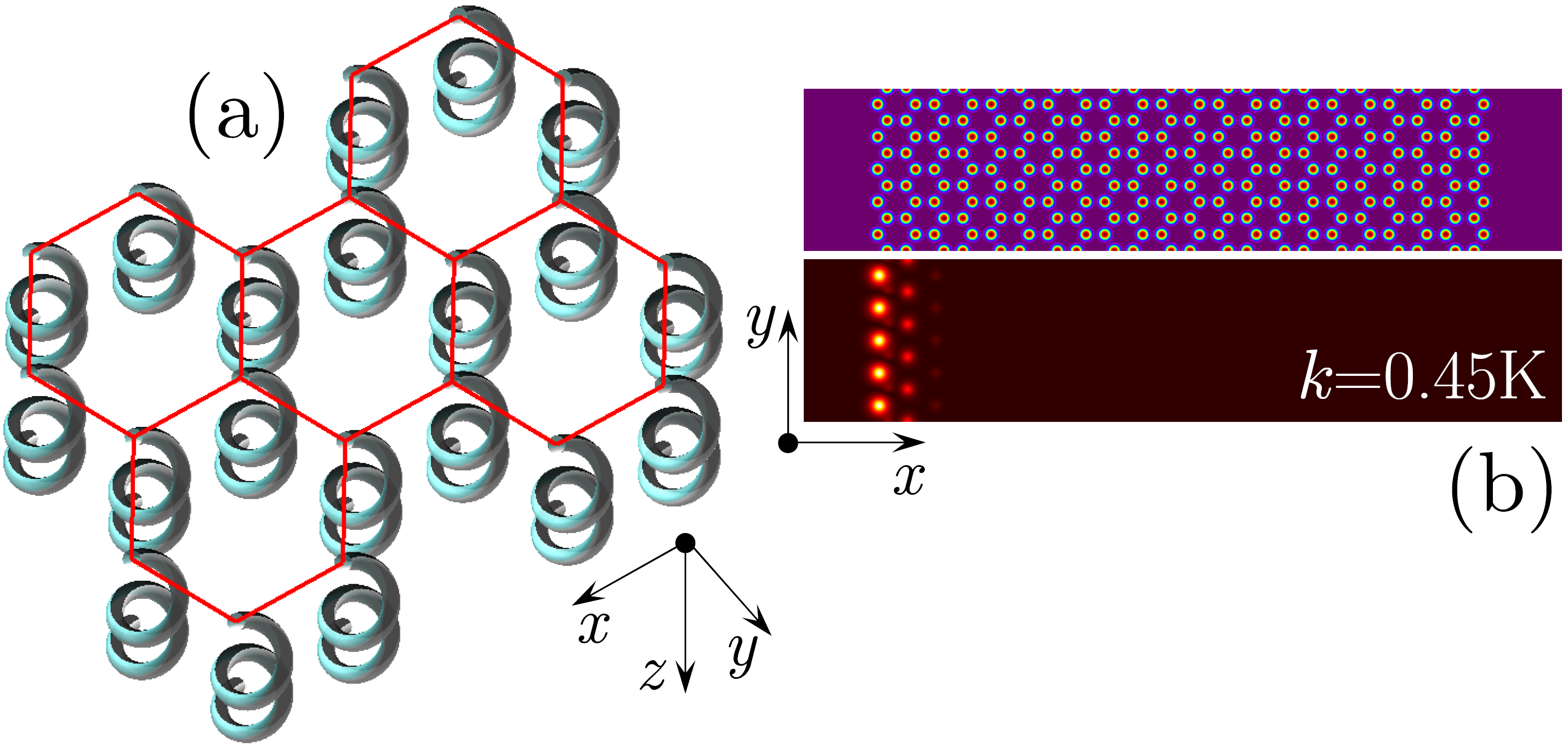}
\caption{(a) Schematics of a 3D  helical honeycomb array with clockwise waveguide rotation direction. (b) Five periods of the truncated array with zigzag-zigzag edges (top) and a linear Floquet edge state  at $k=0.45K$ shown for $z=0$ (bottom). Here and below $p=11$, $d=1.7$, $\sigma=0.4$, $r_0=0.5$, and $Z=8$.}
\label{figure1}
\end{figure}

We consider a two-component field ${\bm E}=(E_x,E_y)$ of the frequency $\omega_0$ propagating along the $z$-axis of helical honeycomb array imprinted in a weakly birefringent medium [Fig.~\ref{figure1}(a)]. Let $\psi_{x,y}  =(L_{d} /L_{s}I_0)^{1/2}e^{\pm i\beta z/4}E_{x,y}$, where $L_{d}= k_x a^2 \approx k_ya^2$ is the diffraction length for close wavenumbers of two components $k_x,k_y$, $\beta=2|k_x-k_y|L_d $ is the normalized phase mismatch, $L_{s} =2c/(\omega_0n_2I_0)$ is the self-action length for nonlinear coefficient $n_2$, $a$ and $I_0$ are the characteristic transverse scale and intensity, respectively. Then, $\psi_{x,y}$ are governed by the system~\cite{Yaroslav}:
\begin{align}  
\label{mainx}
\begin{split}
i\frac{\partial \psi_x}{\partial z}= &\left(-\frac{1}{2} \Delta_\perp   - \frac{\beta}{4}  -  R \right) \psi_x-  \\
&\left( |\psi_x|^2+\frac23|\psi_y|^2 \right) \psi_x-\frac13 \psi_x^*\psi_y^2 ,
\end{split}\\
\label{mainy}
\begin{split}
i\frac{\partial \psi_y}{\partial z}= &\left( -\frac{1}{2} \Delta_\perp  + \frac{\beta}{4} - R \right) \psi_y -  \\
&\left(|\psi_y|^2+\frac23|\psi_x|^2 \right) \psi_y-\frac13 \psi_y^*\psi_x^2 .
\end{split}
\end{align}
Here $\Delta_\perp=\partial_x^2+\partial_y^2$, ${\br}=\textbf{i}x+\textbf{j}y$ and $z$ are normalized to the characteristic transverse scale $a$ and diffraction length $L_d$, respectively. For simplicity, we assume identical refractive index landscapes described by $R(\br,z)$ for both components. The nonlinear terms, in order of their appearance in Eqs.~(\ref{mainx}),~(\ref{mainy}), describe self-phase-modulation (SPM), cross-phase modulation (XPM), and four-wave mixing (FWM).

Honeycomb array [Fig.~\ref{figure1}(a)] is periodic along the $y$ and $z$  axes and is truncated along the $x$-axis: $R(\br+L\textbf{j},z) = R(\br,z+Z) = R(\br,z)$, where $L$  and $Z$ are the respective periods. It is described by the function $R(\br,z)=p\sum_{mn} \exp{\left(-[\br-\br_{nm}-\bs(z)]^2/\sigma^2\right)}$, where $\bs(z) = r_0(\sin(\omega z),\; \cos(\omega z))$ describes helicity of the waveguides, $\br_{mn}$ are the coordinates of the knots of the honeycomb lattice identified by the integers $m$ and $n$, $d$ is the distance between neighbouring waveguides (respectively $L = 3^{1/2}d$), $\sigma$ is the waveguide width, $r_0$ is the helix radius, and $\omega = 2\pi/Z $. We consider truncation creating zigzag-zigzag interface [Fig.~\ref{figure1}(b)]. Such arrays can be fabricated using fs-laser writing~\cite{Rechtsman-13}.

\begin{figure}[t]
\centering\includegraphics[width=\linewidth]{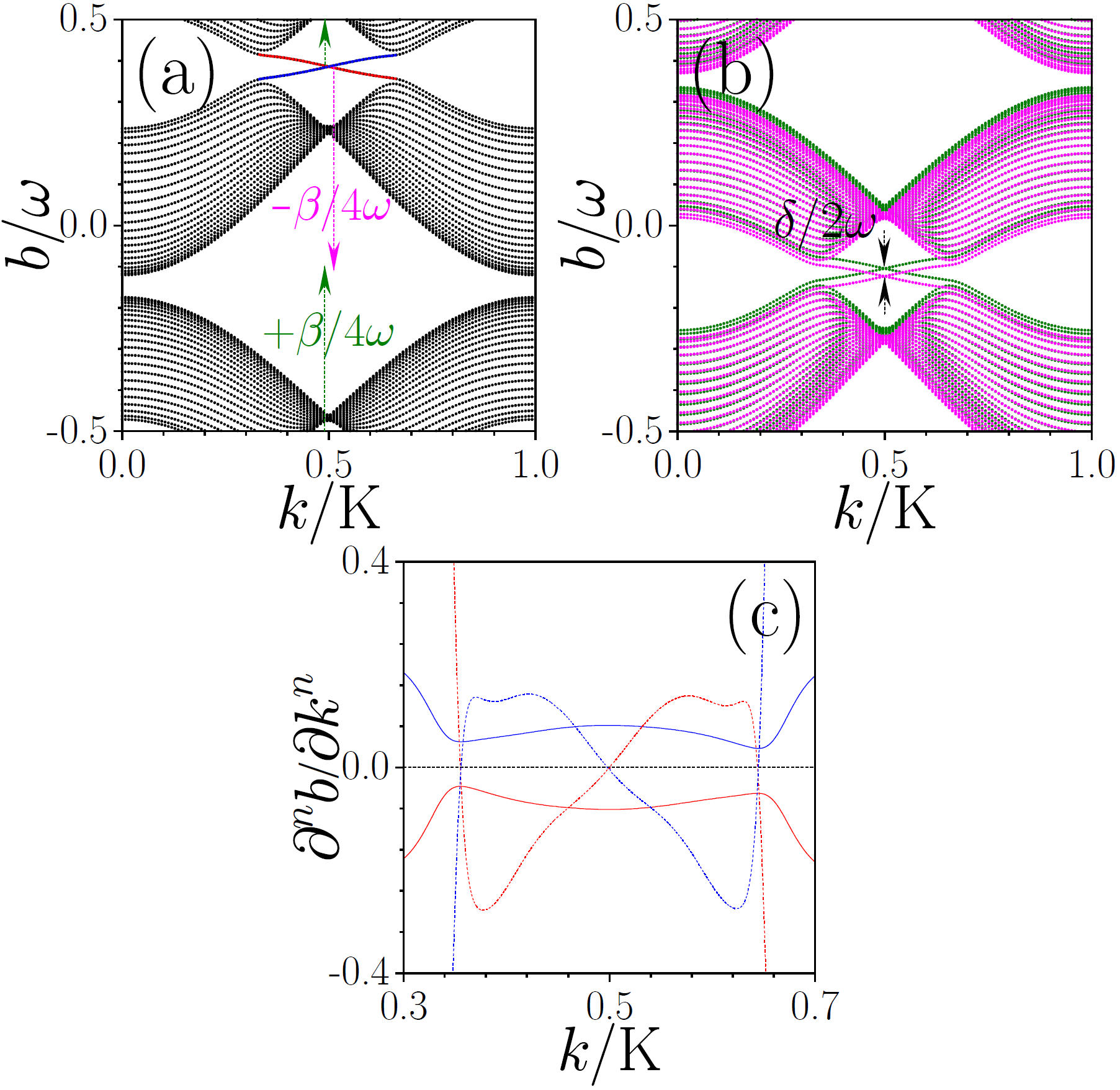}
\caption{(a) Pseudo-propagation constants {\it vs} Bloch wavenumber at $\beta=0$. Green and magenta arrows indicate shifts of the band structures for $\psi_x$ and $\psi_y$ components, respectively, occurring  due to  $\beta>0$ (see the text). (b) Superimposed band structures for $\psi_x$ (green dots) and $\psi_y$ (magenta dots) components at $\beta=1.6$. (c) The first- (solid lines) and second- (dashed lines) order derivatives for the edge state branches  in (a).}
\label{figure2}
\end{figure}

To describe physics of the phenomenon, we start with two linear eigenvalue problems $\left(i\partial_z \pm\beta/4-H\right)\phi_{\nu k}^\alpha=b_{\nu k}^\alpha\phi_{\nu k}^\alpha $, where $H:=-(1/2)\Delta_\perp-R(\br,z)$ and $\alpha$ stands either for $x$  ("$+$") or $y$ ("$-$"). The eigenvalues (pseudo-propagation constants) $b_{\nu k}^\alpha \in [-\omega/2,+\omega/2)$ are marked by the index $\nu$ denoting a band or an edge state, and by the Bloch momentum $k\in[0,\textrm{K})$ with $\textrm{K}=2\pi/L$. At $\beta=0$, the linear spectra $b_{\nu k}^{x}$ and $b_{\nu k}^{y}$ of two components coincide. We illustrate them in Fig.~\ref{figure2}(a), where we omitted indices for brevity. Waveguide helicity leads to opening of a gap with two edge branches between the Dirac points (branches marked by blue and red are localized at the left and right edges, respectively). An example of the edge state from the red branch at $k=0.45K$ is shown in the lower panel of Fig.~\ref{figure1}(b).

Our goal are Floquet solitons bifurcating from the edge states for $x$ and $y$ components in the presence of FWM. To ensure quasi-momentum conservation both components must bifurcate form the linear states with the same Bloch momentum $k_0$. Since, for a given edge there is only one branch of edge modes, we conclude that both linear modes are represented by the same Floquet state with the same indices $(\nu,k_0)$. Below we denote such state by $\phi_0=\phi_{\nu k_0}^xe^{-i(\beta/4)z}=\phi_{\nu k_0}^ye^{i(\beta/4)z}$. At $\beta=0$ the eigenvalue corresponding to $\phi_0$ is $b_0$. Now, to take into account weak nonlinearity and describe the bifurcation of the FWM solitons we look for a solution of ~(\ref{mainx}),~(\ref{mainy}) in the form $\psi_\alpha =\mu e^{i\left(b_0\pm \beta/4\right)z} [ a_\alpha(Y,z)\phi_0+\mu \Phi_\alpha^{(1)} + ...]$, where $\mu\ll 1$ is a formal small parameter, $a_{\alpha}$ are the slowly varying amplitudes and $Y = y - vz$ is the coordinate in the frame moving with velocity $v=-b_0^\prime$ (here $b_0^\prime=\partial b_0/\partial k$). 

Now we recall that the efficiency of the FWM interactions is determined by the phase matching conditions between two components, which would be trivial in $z$-independent media, but acquire the following form in Floquet insulator:
\begin{align}  
\label{b1}
b_{x}=b_{y}+\frac{\delta}{2}, \qquad (\text{mod} \; \omega), 
\end{align}
where $b_{x}=b_0+{\beta}/{4}$,  $b_{y}=b_0-{\beta}/4$, and $|\delta|\ll \omega$ is a small detuning. This condition stems from the fact that the pseudo-propagation constant $b_0$ is determined in the Floquet eigenvalue problem $\left(i\partial_z -H\right)\phi_{0}=b_{0}\phi_{0}$  modulo $\omega$. Therefore (\ref{b1}) is referred below as {\em Floquet phase matching}. Thus, in our Floquet system the exact resonance corresponds to the choice  $\beta = 2n\omega$, $n = 0,\,1,\,2,\,...$, while possible detuning from exact phase matching is given by $\delta =\beta-2n\omega$, where $n$ is chosen to minimize $|\beta - 2n\omega|$. To illustrate the essence of the Floquet phase matching, in Fig.~\ref{figure2}(b) we show the shift of band structures of two components from Fig.~\ref{figure2}(a) arising due to non-zero $\beta$. Since $b_x$ and $b_y$ coincide at $\beta=0$, by adding and subtracting $\beta/4$ one cyclically shifts two band structures upwards [green arrow in Fig.~\ref{figure2}(a)] and downwards [magenta arrow in \ref{figure2}(a)]. If mutual shift equals to $2n\omega$ the displaced spectra exactly coincide with each other [although generally they will not coincide with the original spectrum at $\beta=0$: cf. Fig.~\ref{figure2}(a) and Fig.~\ref{figure2}(b)). In the presence of small detuning $\delta$, shifted Floquet spectra also appear slightly displaced, as shown in Fig.~\ref{figure2}(b). Thus, Floquet phase matching allows to compensate even considerable material phase mismatch $\beta$ leading to efficient FWM interactions.

If condition (\ref{b1}) is satisfied, one can apply the multiscale expansion and $z$-averaging defined by $\langle f \rangle_Z= (1/Z) \int_0^Z f(z)dz$, as described in details in~\cite{IvanovACS-20,IvanovDipol-21}, to derive the equations for the slowly varying envelopes of two components: 
\begin{align}  
\label{Env1}
i\frac{\partial a_x}{\partial z} &= \frac{b_0''}{2}\frac{\partial^2 a_x}{\partial Y^2} - \chi\left(|a_x|^2a_x+\frac23|a_y|^2a_x+\frac13a_x^*a_y^2 e^{-i\delta z} \right), \\
\label{Env2}
i\frac{\partial a_y}{\partial z} &= \frac{b_0''}{2}\frac{\partial^2 a_y}{\partial Y^2} - \chi\left(|a_y|^2a_y+\frac23|a_x|^2a_y+\frac13a_y^*a_x^2 e^{+i\delta z} \right).
\end{align}
Here $b_0''=\partial^2 b_0/\partial k^2$ and $\chi=\langle \int |\phi_0|^4 d\br \rangle_Z$ is the effective nonlinearity coefficient. The derivatives $b_0^\prime=\partial b_0/\partial k$ and $b_0''=\partial^2 b_0/\partial k^2$ versus $k$ for selected states are shown in Fig.~\ref{figure2}(c).

\begin{figure}[t!]
\centering\includegraphics[width=\linewidth]{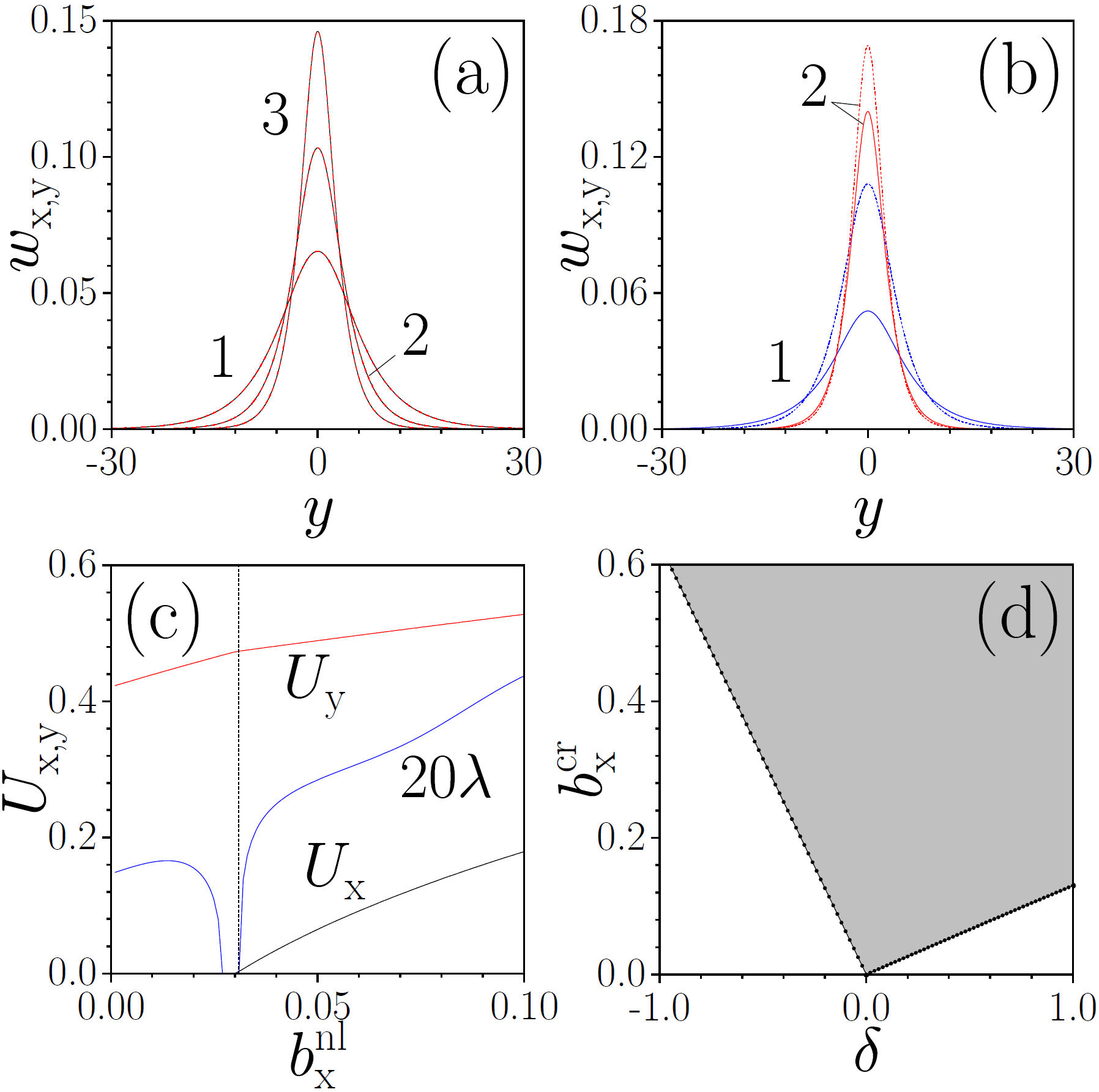}
\caption{Profiles of the $w_x$ (solid lines) and $w_y$ (dashed lines) soliton components at $\delta=0$ (a) and $\delta=0.005$ (b). In (a) $b_x^{nl}=0.002$ (curve 1), $b_x^{nl}=0.005$ (curve 2), and $b_x^{nl}=0.01$ (curve 3). In (b) $b_x^{nl}=0.002$, $b_y^{nl}=0.0045$ (curves 1) and $b_x^{nl}=0.01$, $b_y^{nl}=0.0125$ (curves 2). (c) Powers of components $U_{x,y}$ (black and red curves) and perturbation growth rate $\lambda$ (blue curve) {\it vs} $b_x^{nl}$ at $\delta=0.2$. Vertical dashed line shows the border of existence for vector solitons. (d) Critical propagation constant shift $b_x^{cr}$ {\it vs} detuning $\delta$. All solitons bifurcate from edge state with $k_0=0.45K$, yielding $b_0''=-0.0963$, $\chi=0.703$.}
\label{figure3}
\end{figure}

We are interested in bright vector solitons of Eqs.~(\ref{Env1}),~(\ref{Env2}) that exist at $b_{0}^{\prime\prime}<0$. Such solitons can be found either linearly polarized $a_{x,y} = w_{x,y} \exp{(ib_{x,y}^{nl}z)}$ or elliptically polarized $a_{x} = w_{x} \exp{(ib_{x}^{nl}z)}$, $a_{y} = i w_{y} \exp{(ib_{y}^{nl}z)}$. To ensure that the profiles $w_{x,y}$ are broad and cover many $y$-periods of the structure, the nonlinearity-induced shifts of propagation constants $b^{nl}_{x,y}$ must be much smaller than the pseudo-propagation constants and the width of the gap in the Floquet spectrum. Since Eqs.~(\ref{Env1}),~(\ref{Env2}) allow linearly polarized FWM solutions only for exact Floquet phase matching, $\delta=0$, we consider only elliptically polarized states. Some properties of the FWM Floquet solitons with elliptical polarization are illustrated in Fig.~\ref{figure3}. Representative envelopes of such solitons for exact resonance ($\delta=0$) are shown in Fig.~\ref{figure3}(a). In this case both components coincide and feature equal nonlinear propagation constant shifts $b_x^{nl}=b_y^{nl}$. Examples of profiles at small detuning $\delta \neq 0$ are shown in Fig.~\ref{figure3}(b). Now soliton components have different shapes, since $b_y^{nl}=b_x^{nl}+\delta/2$ is different from $b_x^{nl}$. Such FWM vector solitons bifurcate from the family of scalar solitons with $U_x=0, U_y\neq 0$ at critical propagation constant value $b_x^{nl}=b_x^{cr}$ indicated by the dashed vertical line in Fig. \ref{figure3}(c), the powers $U_{x,y}={\int |\psi_{x,y}|^2d\br}$ of its components monotonically increase with $b_x^{nl}$. Critical propagation constant $b_x^{cr}$ grows with increase of $|\delta|$, so that such vector solitons exist within filled area in Fig.~\ref{figure3}(d) with straight lower borders. We performed linear stability analysis for perturbed solitons $a_{x,y} = (w_{x,y} + u_{x,y}e^{\lambda z} + v_{x,y}e^{-\lambda z})e^{ib_{x,y}^{nl}z}$ of Eqs.~(\ref{Env1}),~(\ref{Env2}), where $ |u_{\alpha}|,\,|v_{\alpha}|\ll|w_{\alpha}|$, and determined growth rate of the most unstable perturbation. While for exact resonance $\delta=0$ all states are stable ($\textrm{Re} \lambda = 0$), for $\delta \neq 0$ we encountered weak oscillatory instabilities with $\textrm{Re} \lambda \neq 0$, see Fig.~\ref{figure3}(c) for dependence of growth rate on $b^{nl}_x$ (blue curve). The characteristic development length for such instabilities is $ 1/(\text{Re}\lambda) \gg Z$. Similar results were obtained for other values of $\delta\neq 0$.

\begin{figure*}[t!]
\centering\includegraphics[width=0.7\linewidth]{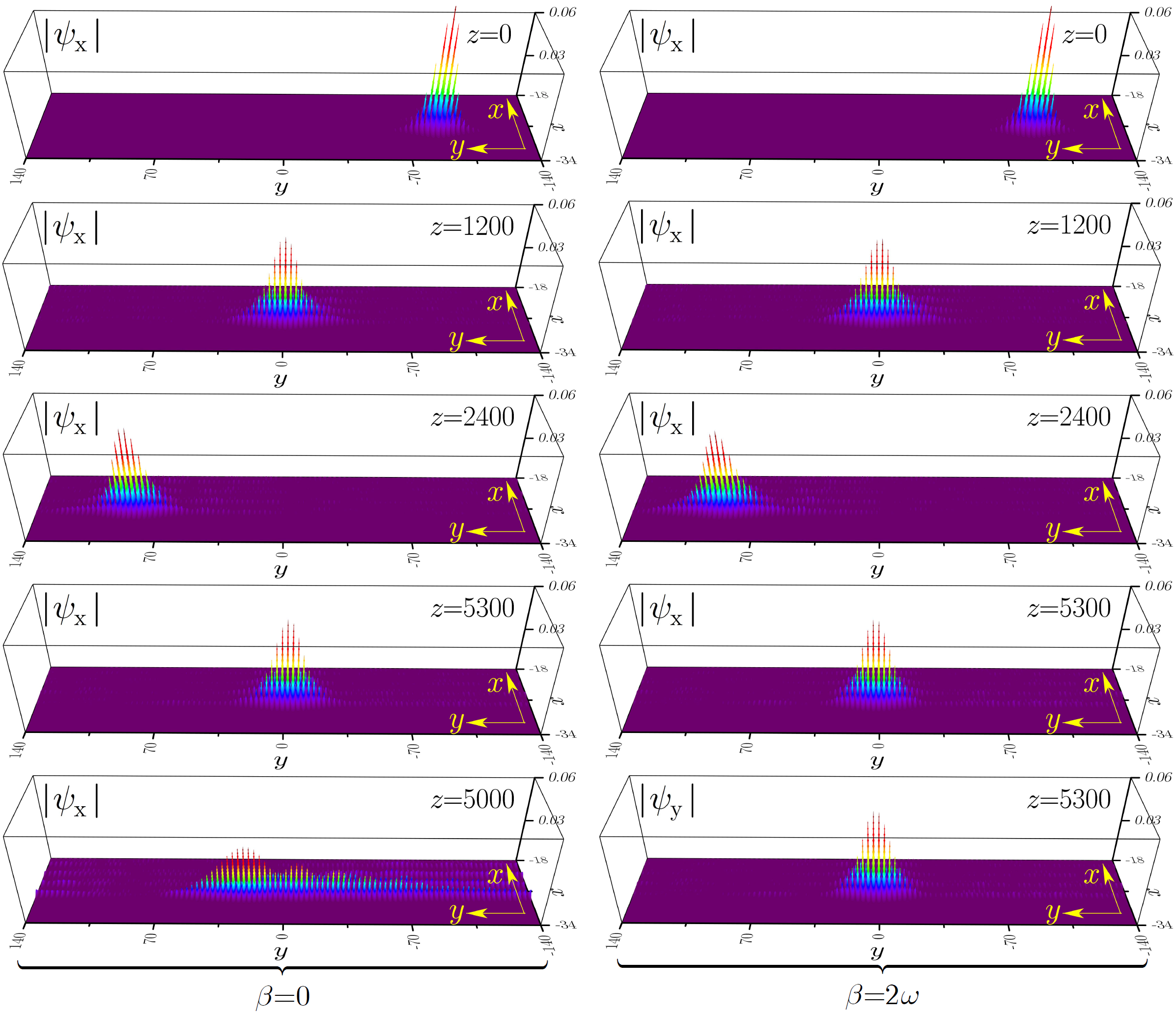}
\caption{Propagation dynamics of the edge soliton corresponding to $b_x^{nl}=b_y^{nl}=0.002$, $k=0.45K$, $b''=-0.0963$, $\chi=0.703$ under conditions of exact Floquet phase matching ($\delta=0$) for $\beta=0$ (left column) and $\beta=2\omega$ (right column). Last panel in the left column shows output wavepacket in the linear regime. Two last panels in the right column compare output $|\psi_x|$ and $|\psi_y|$ distributions.}
\label{figure4}
\end{figure*}

In Fig.~\ref{figure4} we illustrate the evolution of the FWM edge solitons bifurcating from the red branch in Floquet spectrum at $k_0 = 0.45K$ [see Fig.~\ref{figure2}(a)] under conditions of exact Floquet phase matching ($\delta=0$) for $\beta=0$ (left column) and $\beta=2\omega\approx1.57$ (right column). The initial conditions in Eqs.~(\ref{mainx}),~(\ref{mainy}) for these soliton solutions are taken as $a_{\alpha}(y,0)\phi_{0}(\br,0)$ with $a_x$ and $a_y$ found from Eqs.~(\ref{Env1}),~(\ref{Env2}), respectively. We show dynamics for only one component; the dynamics of the second one is identical. Both components are locked and move together along the edge (see two last rows in the right column) without considerable modifications (notice that we work with huge, but finite $y$-window, so due to transverse displacement and periodic boundary conditions in $y$, after sufficiently long propagation distance the soliton reappears from the other side of the window). Weak reshaping, due to the fact that our theory provides $z$-averaged approximation to the exact $z$-oscillating solution, takes place only at the initial stage of propagation. There is almost no radiation into the bulk of the array for selected $r_0$ and $Z$ values. To prove that these states are indeed supported by the combined effect of SPM, XPM, and FWM in the bottom panel of left column of Fig.~\ref{figure4} we show the output distribution propagated through a linear medium. In this case we observe strong asymmetric expansion of the wavepacket. Comparing two columns in Fig.~\ref{figure4} we observe that FWM solitons at two different resonances are very similar [notice that their envelopes obtained from Eqs.~(\ref{Env1}),~(\ref{Env2}) are identical, but they were propagated using different resonant values of $\beta$ in full 2D Eqs.~(\ref{mainx}),~(\ref{mainy})].

\begin{figure}[t!]
\centering\includegraphics[width=0.95\linewidth]{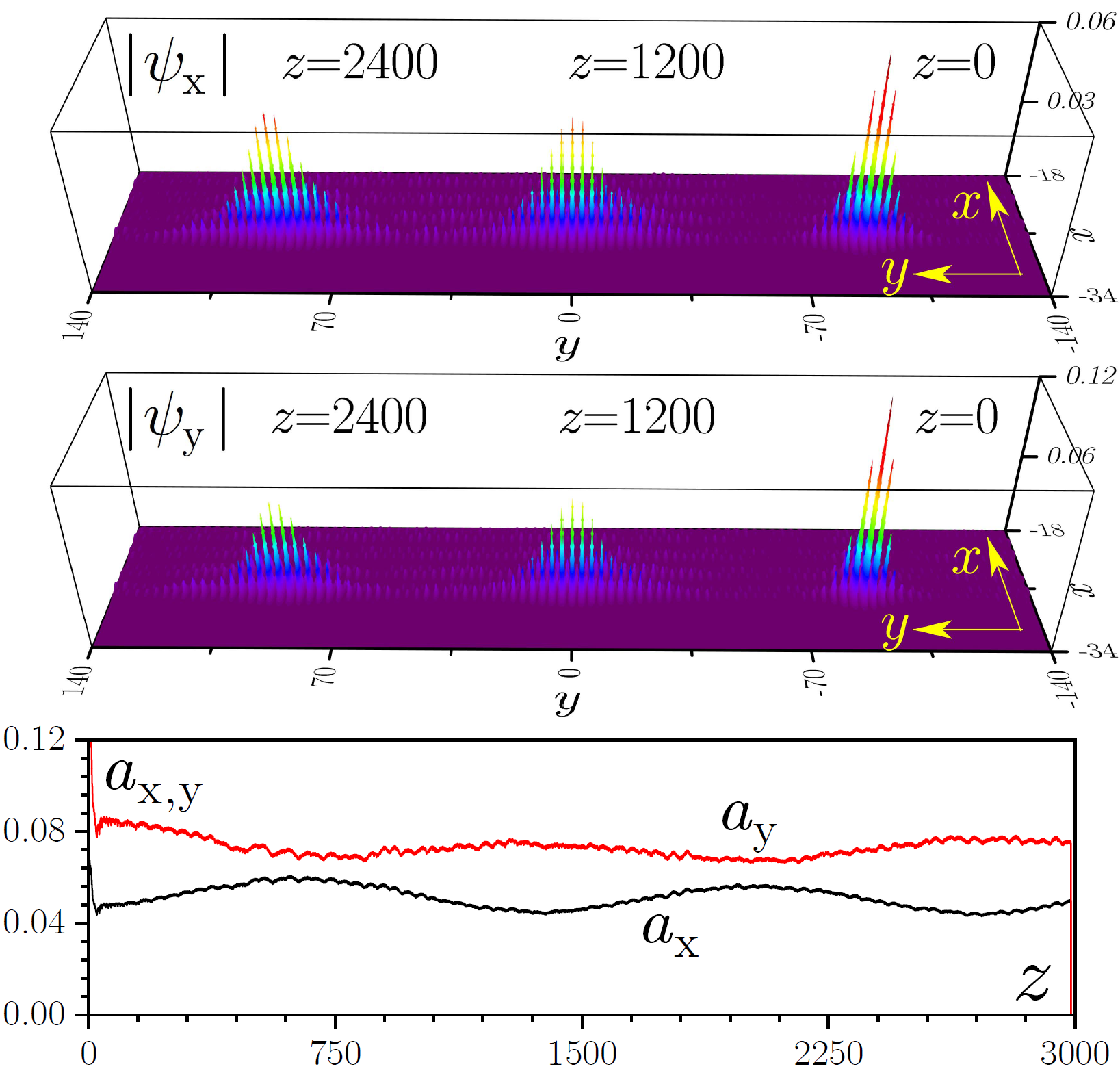}
\caption{Propagation dynamics of the edge soliton corresponding to $\delta=0.005$, $b_x^{nl}=0.002, b_y^{nl}=0.0045$, $b_0''=-0.0963$, and $\chi=0.703$ at $k=0.45K$ (top and middle). $|\psi_x|$ and $|\psi_y|$ distributions at different distances are superimposed on the same plot.   Peak amplitudes of components versus $z$ (bottom).}
\label{figure5}
\end{figure}

Propagation in the frames of Eqs.~(\ref{mainx}),~(\ref{mainy}) also confirms metastability of the vector solitons at $\delta=0.005$, which survive over hundreds of helix periods. The evolution of $\psi_x$ and $\psi_y$ components of such soliton is illustrated in top and middle panels of Fig.~\ref{figure5}, where distributions at different distances $z$ are superimposed on the same plot. Metastability leads to small-amplitude oscillations (bottom row) that do not cause destruction of this state at the considered distances up to $z \sim 10^4$, even though linear stability analysis predicts formal instability of the envelope and small growth rate $\lambda=4.3\cdot 10^{-4}$.

In conclusion, we have shown that in a Floquet topological system it is possible to realise a new type of resonant phase matching that enables the existence of a new type of FWM edge solitons that are stable at exact resonance and become metastable at small detuning from resonance.

\medskip

The authors acknowledge funding from the RSF (grant  21-12-00096),
and Portuguese Foundation for Science and Technology (FCT) under Contract no. UIDB/00618/2020.

\medskip

\noindent\textbf{Disclosures.} The authors declare no conflicts of interest.


\end{document}